\documentclass[12pt]{article}
\usepackage{a4}
\usepackage{epsf}

\begin{document}

\newcommand{\p}[1]{\frac{\partial}{\partial#1}}
\newcommand{\pd}[1]{\frac{\partial^2}{\partial#1^2}}
\newcommand{\ket}[1]{\left|#1\right>}
\newcommand{\state}[1]{\left|#1\right>}
\newcommand{\ap}{\alpha'}

\newcommand{\problem}[1]{{\bf !!}{\em #1}{\bf !!}}

\newcommand{\m}[1]{\mathrm{#1}}

\begin{center}
\vskip 30pt
{\bf\Large 0-brane Quantum Chemistry}
\vskip 50pt
\small{V\'ACLAV KARE\v{S}\footnote{\normalsize{\emph{E-mail:
kares@physics.muni.cz}}} \\ 
\emph{Institute for Theoretical Physics, Masaryk University,\\
 Kotl\'a\v{r}sk\'a 2, 611 37, Brno, Czech republic}}
\vskip 20pt
\end{center}

\begin{abstract}
We apply two different numerical methods to solve for the
boundstate of two 0-branes in three dimensions. One method is developed
by us in this work and we compare it to a method existing in the
literature. In spite of considering only three dimensional
Minkowski space we obtain interesting results which should
give some basic understanding of the behaviour of 0-branes.

\end{abstract}

\vskip 50pt

\section{Introduction}

D-branes are very important non-perturbative objects in string theory.
Their existence is essential since they are needed, among other things,
for various string theory dualities to work. The most primitive definition 
of a D-brane is as a hyper surface on which open strings end.

In this work we are interested only in D0-branes. They
play a particularly important role since they are the basic constituents
of Matrix theory \cite{BS}, which, being a suggestion for a non-perturbative
definition of string theory (or M-theory \cite{BFSS}), deserves particular
attention. It is also possible to think about them as bound states of higher
dimensional unstable D-branes \cite{H,W}. Furthermore, as was shown by Myers
\cite{M}, they can also form bound states with all the properties of higher
dimensional branes. It is therefore clear that D0-branes have many
interesting properties that make them worth studying in
detail. Concretely this means studying supersymmetric quantum mechanics 
which is an interesting topic in itself. Although this topic has been
studied before \cite{W1}-\cite{SK} we feel that there are still unresolved
issues. In particular one could use a computer to compute properties of
D0-branes which are not possible to address analytically because of
the complexity of the theory. This is the goal of this work, to
develop a ``0-brane Quantum Chemistry''. It should be mentioned that
similar issues have been addressed in \cite{W1}. However, we use a
different method which we compare to the method developed in \cite{W1}
(to which we also suggest certain improvements). 

More concretely, in this work we try to find the bound state of two
D0-branes in three dimensional Minkowski space (which is really a toy-model
for the real situation, D0-branes in ten dimensional Minkowski space). 
It should also be noticed that the really interesting cases where the
D0-brane theory is thought to describe macroscopic supergravity states
is achieved by taking the number of zero branes to infinity.

\section{The model}
\label{model}

The low energy physics of $N$ parallel D$p$-branes is governed by the
dimensional 
reduction of $9+1$ dimensional ${\cal N}=1$ supersymmetric Yang-Mills theory
with $U(N)$ gauge group to $p+1$ dimension \cite{Der}. The center of mass 
motion is governed by the overall $U(1)$ factor so if we are interested in 
relative motion only, we can choose the gauge group to be $SU(N)$. In the 
case we are interested in, the relative motion of two 0-branes, we thus 
choose the gauge group to be $SU(2)$. The action is \cite{KP}
\begin{eqnarray}S&=&\int\mathrm{d}t\bigg[\frac{1}{2g_s}\dot 
X^{a}_{i}\dot X^{a}_{i}+  \frac{i}{2}\psi^{a}_A\dot \psi^{a}_A-
  \frac{1}{4g_s}\left(\epsilon^{abc}X^{b}_{i}X^{c}_{j}\right)^{2}+
  \frac{i}{2}\epsilon^{abc}X^{a}_{i}\psi^{b}_A(\gamma^{i})_{AB}
  \psi^{c}_B \nonumber \\
&&+\frac{1}{g_s}\epsilon^{abc}\dot X^{a}_{i}A^{b}_{0}X^{c}_{i}+
  \frac{1}{2g_s}\left(\epsilon^{abc}A^{b}_{0}X^{c}_{i}\right)^{2}-
  \frac{i}{2}\epsilon^{abc}A^{a}_{0}\psi^{b}_A\psi^{c}_A\bigg]
\label{full}
\end{eqnarray}
and the Hamiltonian derived from this is \cite{KP}
\begin{eqnarray}
H&=&\frac{g_s}{2}\left(\pi^{a}_{i}\right)^{2}+
    \frac{1}{4g_s}\left(\epsilon^{abc}X^{b}_{i}X^{c}_{j}\right)^{2}-
    \sum_{i=1}^{7}\epsilon^{abc}X^{a}_{i}\bar\chi^{b}_A(\tilde\gamma^{i})_{AB}
    \chi^{c}_B\nonumber \\
 &&-\frac{1}{2}\epsilon^{abc}X^{a}_{8}\left(\chi^{b}_A\chi^{c}_A-
    \bar\chi^{b}_A\bar\chi^{c}_A\right)-
    \frac{i}{2}\epsilon^{abc}X^{a}_{9}\left(\chi^{b}_A\chi^{c}_A+
    \bar\chi^{b}_A\bar\chi^{c}_A\right)
\label{hamiltonian}
\end{eqnarray}
together with the constraint one gets from varying (\ref{full}) with 
respect to $A_0$
\begin{eqnarray}
G^a\equiv\epsilon^{abc}(X_i^b\pi_i^c-i\bar\chi_A^b\chi_A^c)=0\;.
\label{gen}
\end{eqnarray}
On quantum level, we restcrict our Hilbert space to vectors which satisfy
\begin{eqnarray}
G^a\state{\Psi}=0
\label{phys}
\end{eqnarray}
that is, our physical space is gauge invariant because $G^a$ are gauge
generators.

We will only study motion of two $0$-branes in three dimensional
Minkowski space. We hope that this gives us the basic behavior of $0$-branes
and also an understanding of the full problem. This problem of two branes
is also described in \cite{KP} but we study it in a different 
way. In the three dimensional case the action is given by 
dimensional reduction of ${\cal N}=1$ supersymmetric Yang-Mills theory 
with $SU(2)$ gauge group in $2+1$ dimension to $0+1$ dimension. 
The Hamiltonian takes the slightly simpler form
\begin{eqnarray}
H&=&\frac{g_s}{2}\left(\pi^a_i\right)^2+
  \frac{1}{4g_s}\left(\epsilon^{abc}X^b_1X^c_2\right)^2\nonumber\\
&&-\frac{1}{2}\epsilon^{abc}X^a_1\left(\chi^b\chi^c-
   \bar\chi^b\bar\chi^c\right)-
\frac{i}{2}\epsilon^{abc}X^a_2\left(\chi^b\chi^c+
   \bar\chi^b\bar\chi^c\right)
\label{toy}
\end{eqnarray}
where $X_i$ are fields in the $SU(2)$ adjoint representation and $\chi$ 
is a complex fermion also in the adjoint representation. Of course we 
still have to impose gauge invariance (\ref{phys}). In fact, the gauge 
invariance complicates things somewhat since we would like to separate 
out gauge invariant degrees of freedom from pure gauge degrees of 
freedom in our basic quantum mechanical operators $X^a_i$ and 
$\pi^a_i$. Let us focus on physical content of the $X_i^a$. It contains 
six components (the gauge index runs over three values and the space index 
$i$ runs from 1 to 2). We know that we can remove three of these variables 
using gauge transformations so only three variables are observable. These 
three variables should describe the relative position of two pointlike 
objects in two space dimensions. We draw the conclusion that one of the 
physical variables do not have the interpretation of a coordinate but 
rather as some internal auxilliary degree of freedom.

To get some further
insight into this problem it is neccesery to investigate the bosonic 
vacuum of the theory. It is possible to explicitly separate the gauge 
degrees of freedom from $X_i^{~a}$ by decomposition it in matrix form 
\cite{HS}
\begin{eqnarray}(X)_{ai}=(\psi)_{ar}(\Lambda)_{rs}(\eta)_{si} \;.
\label{decomposition}
\end{eqnarray}
Here the matrix $\psi$ is an group element in the adjoint representation 
of $SU(2)$. Thus when the gauge group acts on $X^a_i$, $\psi$ just changes 
by ordinary gauge group multiplication (from the left). We will 
parametrise the group element $\psi$ by the "angles" $\alpha,\beta,\gamma$ 
(\ref{psi}). This decomposition has the advantage that all the gauge 
dependence sits in $\psi$ and all the other matrices are gauge invariant. 

In an analogous way we have separated out the dependence on rotations in 
space. Namely, performing an $SO(2)$ rotation in space we have an element 
of $SO(2)$ acting from the right on the matrix $X_{ai}$. Thus we can 
separate out the dependence on the angle in space (we will call it 
$\phi$) by saying that $\eta$ is a group element of $SO(2)$.

\begin{figure}[htb]
\begin{center}
\mbox{\epsfxsize=6.5cm\epsfbox{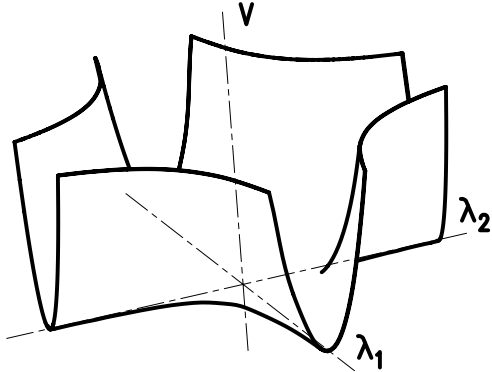}}
\end{center}
\caption{Bosonic potential}
\label{picpot}
\end{figure}

We are left with the matrix $\Lambda$ (\ref{rt}) which by construction 
is both gauge and space rotation invariant
\begin{eqnarray}
\Lambda&=&\left(\matrix{\lambda_1 & 0 \cr 0 
& \lambda_2  \cr0 &0}\right)\;.
\end{eqnarray}
The bosonic potential in
 (\ref{toy}) is gauge and rotation
invariant and in the new decomposition coordinates depends only on two
$\lambda_i$ which have length dimension (Fig. \ref{picpot}).

The parametrisation
\begin{eqnarray}
\lambda_1=r\cos\theta;\;\lambda_2=r\sin\theta
\label{param}
\end{eqnarray}
is the only way how to obtain exactly one variable, $r$, with the 
dimension length, which could represent relative distance of two branes. 
The dimensionless $\theta$ is the auxilliary coordinate. The potential in 
this coordinate reads
\begin{eqnarray}
\frac{1}{8g_s}r^4\sin^22\theta \;.
\label{potential}
\end{eqnarray}

Looking at the picture we can draw some interesting conclusions.
If we fix a point on the bosonic vacuum (a classical static
configuration with minimum energy), that is on the axes, we can study the
behavior of the potential for small fluctuations of the auxilliary 
variable $\theta$. We see that for large $r$ the $\theta$ fluctuation are 
very much suppressed but at small $r$, $\theta$ will be allowed to 
fluctuate. This can be interpreted to mean that when the branes get close 
to each other, they can start to move also in the $\theta$ direction. Thus,
$\theta$ is an auxilliary coordinate which is visible only
when the branes come close together.

The above discussion included only the bosonic degrees of freedom, we
should keep in mind that the fermionic degrees of freedom can (and
will) change this behavior somewhat. In essence, the Pauli repulsion
will try to spread out the wavefunction as much as possible.

 When we use the standard operator representation of $\pi_i^a$ and
$X^a_i$ the first term in the Hamiltonian (\ref{toy}) is proportional to
the Laplacian which we have to rewrite in the decomposition coordinates
above to separate out gauge using formula
\begin{eqnarray}
\frac{1}{\sqrt g}\partial_i(\sqrt g g^{ij}\partial_j) 
\label{formula}
\end{eqnarray}
where the $g$ is the metric which is trivial in $X^a_i$ coordinates. It is
also good idea to rewrite the Laplacian
in terms of gauge angular momenta $L^a\equiv\epsilon^{abc}X_i^b\pi^c_i$:
\begin{eqnarray}
L^1&=&-i\cot\gamma\sin\alpha\p{\alpha}+i\csc\gamma\sin\alpha\p{\beta}
      +i\cos\alpha\p{\gamma} \nonumber \\
L^2&=&-i\cot\gamma\cos\alpha\p{\alpha}+i\csc\gamma\cos\alpha\p{\beta}
      -i\sin\alpha\p{\gamma} \nonumber \\
L^3&=&i\p{\alpha} \;,
\label{gaugemomenta}
\end{eqnarray}
and physical angular momentum
\begin{eqnarray}
L^0\equiv\epsilon^{ij}X_i^a\pi_j^a=-i\p\phi
\label{momentum}
\end{eqnarray}
since their action on states with given gauge and rotational properties 
is simple. In particular, the (bosonic) ground state should have total 
spin equal to zero and be gauge invariant. All angular momenta are Killing 
vectors of the metric. 

In Appendix \ref{applaplace} we derive the lagrangian expressed in terms 
of gauge invariant variables and angular momenta
\begin{eqnarray}
L=\frac{1}{r^5}\p{r}r^5\p{r}+\frac{1}{r^2\sin4\theta}
\p{\theta}\sin4\theta
\p{\theta}-({\bf\Psi\Pi}^{-1}{\bf\Psi}^{-1})_{\mu\nu}L^\mu L^\nu \;.
\label{laplacian}
\end{eqnarray}
Here $L^\mu$ $\mu = 0,1,2,3$ is a compact notation for the 
(physical and gauge) angular momenta defined above. Furthermore, we have 
defined
\begin{eqnarray}
{\bf\Psi}=\left(\matrix{-1 & \cr & \psi}\right)
\label{boldpsi}
\end{eqnarray}
and $\bf\Pi$
\begin{eqnarray}
\left(\matrix{r^2 & & & r^2\sin 2\theta \cr
 & r^2\sin^2\theta & & \cr
&  & r^2\cos^2\theta & \cr
r^2\sin 2\theta & & & r^2}\right) \;.
\label{boldpi}
\end{eqnarray}

To find the (bosonic) ground state we will find all gauge invariant states 
with spin zero. The first state is the vacuum state $\state{0}$. Then
we may act with the fermionic creation operators $\chi^a$ on the vacuum 
to find new states. The following states has total spin zero and they are 
gauge invariant
\begin{eqnarray}
\state{r}=\frac{1}{2}\psi_{ar}\state{a}=
\frac{1}{2}e^{\m{i}\phi}\psi_{ar}\epsilon^{abc}\chi^b\chi^c\state{0}
\;.
\label{states}
\end{eqnarray} 
That is, they satisfy
\begin{eqnarray}
G^a\state{r}=0\;.
\end{eqnarray}
The most general gauge invariant wavefunction with
total spin zero can then be written
\begin{eqnarray}
g(r,\theta)\state{0}+f_r(r,\theta)\state{r} \;.
\label{gaugestate}
\end{eqnarray}
It would also possible to construct the superpartner ground state with 
the help of
\begin{eqnarray}
&&\state{r'}=e^{i\phi}\psi_{ar}\chi^a\state{0}\nonumber\\
&&\state{0'}=\frac{1}{6}e^{2i\phi}
             \epsilon^{abc}\chi^a\chi^b\chi^c\state{0} \;.
\end{eqnarray}

Now we study what happens when we
act with the Laplacian (\ref{laplacian}) on this wavefunction and using the
result we write the Hamiltonian matrix in this base. Let us start with second
part of the Laplacian which is the relevant bosonic piece
\begin{eqnarray}
-({\bf\Psi\Pi}^{-1}{\bf\Psi}^{-1})_{ab}L^aL^b\state{r}&=&
\state{s}{\bf\Pi}^{-1}_{sr}-\state{r}\mathrm{Tr}{\bf\Pi}^{-1}\nonumber \\
-2({\bf\Psi\Pi}^{-1}{\bf\Psi}^{-1})_{0a}LL^a\state{r}&=&
2\mathrm{i}\state{u}{\bf\Pi}^{-1}_{0s}\epsilon_{usr}\nonumber \\
-({\bf\Psi\Pi}^{-1}{\bf\Psi}^{-1})_{00}LL\state{r}&=&
-\state{r}{\bf\Pi}^{-1}_{00}
\end{eqnarray}
where $\mathrm{Tr}$ is only on the gauge indexes. Finally the fermionic
interaction in the Hamiltonian gives us
\begin{eqnarray}
H_F\state{r}&=&-\state{0}(\Lambda\eta)_{r1}+
\mathrm{i}\state{0}(\Lambda\eta)_{r2}\nonumber\\
H_F\state{0}&=&-\state{r}(\Lambda\eta)_{r1}-
\mathrm{i}\state{r}(\Lambda\eta)_{r2}\;.
\end{eqnarray}
So the full Hamiltonian matrix elements are
\begin{eqnarray}
H_{rs}&=&-\frac{g_s}{2}\left(h+{\bf\Pi}^{-1}_{rs}-\delta_{rs}\mathrm{Tr}
{\bf\Pi}^{-1}+2\mathrm{i}{\bf\Pi}^{-1}_{0u}\epsilon^{rus}-\delta_{rs}
{\bf\Pi}^{-1}_{00}\right)+
\frac{1}{8g_s}r^4\sin^22\theta\nonumber\\
H_{0r}&=&-(\Lambda\eta)_{r1}+\mathrm{i}(\Lambda\eta)_{r2} \nonumber \\
H_{00}&=&-\frac{g_s}{2}h+\frac{1}{8g_s}r^4\sin^22\theta
\end{eqnarray} 
where
\begin{eqnarray}
h=\frac{1}{r^5}\p{r}r^5\p{r}+
\frac{1}{r^2\sin4\theta}\p{\theta}\sin4\theta\p{\theta}
\;.
\end{eqnarray}
Notice that the Hamiltonian we have obtained is the same as in \cite{SK} but
differs from the one used in \cite{KP}.

\section{Numerical calculation I}

This section gives an overview of solving the Hamiltonian eigenvalue 
problem (\ref{toy}) using the numerical renormalized Numerov method 
\cite{J}. With it one can solve for the discrete spectra of a one 
dimensional operator of the form
\begin{eqnarray}
-\frac{1}{2}\pd{x}{\bf 1}+{\bf V}(x)
\label{op}
\end{eqnarray}
which acts on $L[a,b]\otimes{\bf C}^n$ with Dirichlet boundary
conditions. However, our Hamiltonian depends on the {\em two} coordinates
$r$ and $\theta$ (the other coordinates, the angles, are fixed by the
requirement that we are studying only gauge invariant states with
spin zero). The ``kinetic'' term in our Hamiltonian (i.e. the term which
contains the derivatives) is
\begin{eqnarray}
\frac{1}{r^5}\p{r}r^5\p{r}+
\frac{1}{r^2\sin4\theta}\p{\theta}\sin4\theta\p{\theta}
\label{kinetic}
\end{eqnarray}
and we see that it is naively not of the form required above.
Let us sketch briefly how to modify our problem to be able to apply the
method. An arbitrary wavefunction can be written in the form
\begin{eqnarray}
\ket{\Psi}=\sum_{i,j} \Psi_{ij}(r)Y_{ij}(\theta)e_i
\label{state}
\end{eqnarray}
where $\{Y_{ij}(\theta),j=1,\dots\}$ is a complete basis of functions
(with apropriate boundary conditions) in $\theta$
(the explicit choice of basis does not have to be the same for
different values of the index $i$ but can be chosen to optimize the numerics).
The functions $\{\Psi_{ij},j=1,\dots\}$ (which we will compute by the
Numerov method) one can 
think of as being ``combination coefficients'' depending continuously
on $r$.
$\{e_i\}$ is the standard base spanning ${\bf C}^n$, in our case it is
the four dimensional complex vector space on which the Hamiltonian
matrix acts.
Choosing a concrete basis depends only on the Hamiltonian domain which
depends on one of the 
selfconjugated extensions of the Hilbert space (an extension of the
Hilbert space such that the Hamiltonian operator is hermitian)
but we will rather apply a physical principle which will be described
later. The expectation values 
of the Hamiltonian in the base
\begin{eqnarray}
Y_{ij}e_i\;\emph{(no sum)}
\label{base}
\end{eqnarray}
gives the same number of coupled equations for the radial part as the number of
$\Psi_{ij}$ in (\ref{state}). Thus the problem is now correctly defined and
the matrix representation of the expectation value of the Hamiltonian
in the basis above forms the potential which is used in the one dimensional Hamiltonian (\ref{op}). To get ${\bf 1}$ in
the kinetic term we have to orthonormalize the base. Furthermore, to
be able to use the Numerov method on a computer we need a finite basis
which means that we need to ``cut off'' or restrict the base to be finite. 
The rescaling of the wavefunction
$\ket{\Psi}$ by the factor $r^{5/2}$ transforms the radial part
of (\ref{kinetic}) to the operator
\begin{eqnarray}
\pd{r}-\frac{15}{4}r^2 
\end{eqnarray}
and changes boundary condition to Dirichlet.

  Let us now describe the physical principle we use to choose the
basis functions $Y_{ij}(\theta)$
for our Hamiltonian (\ref{toy}).

 We define the wavefunction on the
interval $\theta\in[0,\pi/4]$ (\ref{domain}). It is not necessary to choose
this particular interval, one could, for instance, select the interval
$[-\pi/4,0]$ instead of the above mentioned.  Using the identification
$\tilde\theta=-\theta$ the Hamiltonian  defined on the interval
$\tilde\theta\in[-\pi/4,0]$ acting on states with total spin
zero is connected to our original Hamiltonian on the
interval $\theta\in[0,\pi/4]$ by the unitary transformation
\begin{eqnarray}
H(\tilde\theta)=U^\dagger H(\theta)U
\label{unitary}
\end{eqnarray}
where
\begin{eqnarray}
U=\left(\matrix{1 &   &   &  \cr
                  & 1 &   &  \cr
		  &   & -1&  \cr
		  &   &   &  1}\right).
\end{eqnarray}
It is also possible to consider other interval $\tilde\theta\in[\pi/4,\pi/2]$.
Using the identification $\tilde\theta=\pi/2-\theta$ the corresponding
Hamiltonian can be also obtained by a unitary transformation with the 
matrix
\begin{eqnarray}
U=\left(\matrix{1 &   &   &  \cr
                  &  &  -\mathrm{i} &  \cr
		  &  \mathrm{i} & &  \cr
		  &   &   &  1}\right).
\end{eqnarray}
The wavefunctions of course also transform under the unitary 
transformation
\begin{eqnarray}
\ket{\Psi(\theta)}=U|\Psi(\tilde\theta\left)\right> \;.
\end{eqnarray}
If we require that the wavefunctions be everywhere smooth, the above 
condition severely restricts the possible wavefunctions and in particular 
the basis wavefunctions that we can use.
We note that this principle was not used in \cite{KP} and hence the 
$\theta$ derivative of their groundstate wavefunction at $\theta = \pi/4$ 
is not well defined.

For the bosonic part of the wavefunction we will use the basis states
\begin{eqnarray}
  \cos (4j\theta) e_1 \quad j=0,\dots
\label{boson}
\end{eqnarray}
and for the fermionic part we will use
\begin{eqnarray}
-\mathrm{i}\cos\theta\cos (2j\theta) e_2
+(-1)^j\sin\theta\cos (2j\theta) e_3 \quad j=0,\dots
\label{fermion}
\end{eqnarray}
They nicely cancel the
divergences in the 
Hamiltonian coming from the Laplacian.
Of course these basis functions satisfy
the boundary condition above and they also form a complete basis for the
functions with the given boundary conditions. However, the basis is not
orthonormal due to the non trivial $\theta$ part in the measure (\ref{measure}).
So we need to orthonormalize these basis functions to get ${\bf 1}$ in
(\ref{op}). Doing so, using the Gramm-Schmidt procedure, we diagonalize the
$\theta$ part of the Laplacian.
Notice that these two sets of basis functions also correspond to the
bases which are implicitly used in the Wosiek method \cite{W1} when one
are doing calculations with states with total
spin zero (the basis functions above depend on the total spin).
 This will be shown in  section
\ref{comparison}. Thus we will be able to compare our results and our method very
directly with the results obtained by the Wosiek method \cite{W1}.

  It is not obvious how much our results for a fixed number of basis
functions fit the exact solutions which one would get using the
complete basis. To get some intuition for how the general solution would
look like we will repeat the calculations increasing the number of
basis functions each time and hopefully one can extrapolate the result
to the exact case. At least we should be able to make an intelligent
guess at the properties of the exact solution.

To study the groundstate of our Hamiltonian (which, because of
supersymmetry, should have energy zero \cite{N}) is a good test for
the method described above.

 There are some  results for coupling constant
$g_s=0.1$ in the table \ref{tab} (p. \pageref{tab}) where $N$ is the
number of the test functions (\ref{boson},\ref{fermion}) and $E$ is their
corresponding groundstate energy. The dependence of the energy on the number of
functions we include in the basis given in the
last two columns of the table
 is as follows
\begin{eqnarray}
E=\frac{1.44}{N}\;.
\label{dep}
\end{eqnarray}
As was claimed in \cite{W1}, one can therefore predict the groundstate energy
as a function of the number of basis functions with very high
accuracy. We therefore see that in any concrete numerical calculation
(using a finite basis) we do not expect to get zero energy. Only in
the (numerically unobtainable) case of infinite basis do we get zero energy.

\begin{figure}
\begin{center}
\mbox{\epsfbox{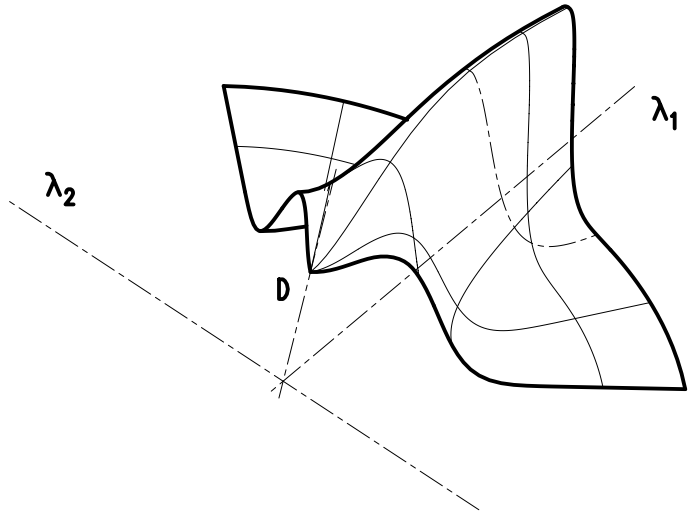}}
\end{center}
\caption{Probability density}
\label{picden}
\end{figure}

The picture (Fig. \ref{picden}) is the probability density for the case with the 
highest number of basis functions in the table.
The following two pictures show the bosonic contribution coming 
from (\ref{boson}) and the fermionic one coming from (\ref{fermion}).
\begin{figure}
\begin{center}
\mbox{\epsfbox{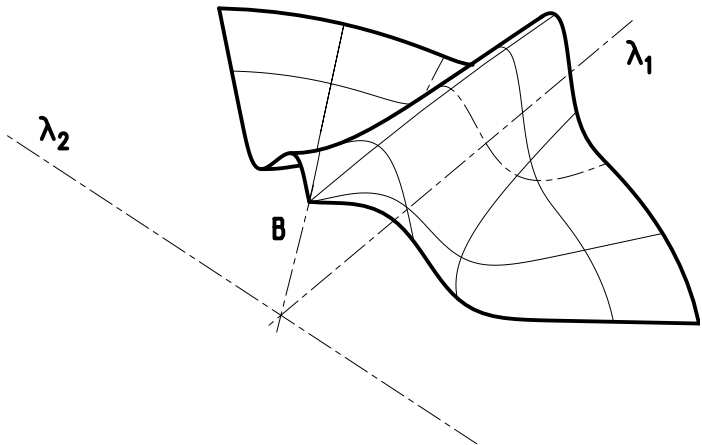}}
\end{center}
\caption{Bosonic part of the probability density}
\label{bosprop}
\end{figure}
\begin{figure}
\begin{center}
\mbox{\epsfbox{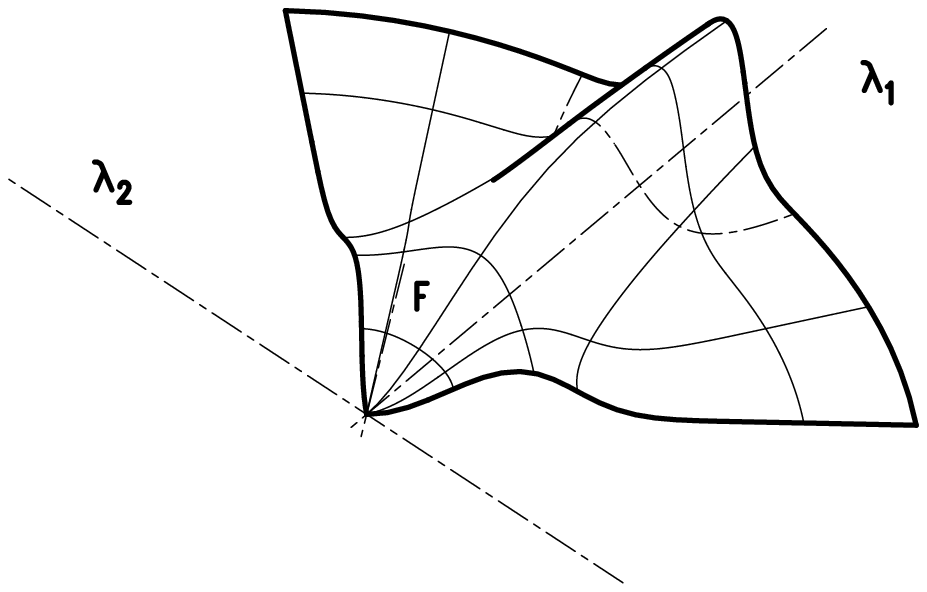}}
\end{center}
\caption{Fermionic part of the probability density}
\label{ferprop}
\end{figure}
The domain of these plots is
$(r,\theta)\in[0,2.2]\times[-\pi/4,\pi/4]$. The hill of the
probability density is located at the boson potential valley (\ref{potential})
and isolines represents sections for fixed $r$, $\theta$ and one for fixed
density on each picture. The maximum of the probability density of any
constant $r$ section is in the potential valley ($\theta=0$). Notice that the global
maximum is not at $r=0$. This is probably an effect of the fermion 
Pauli repulsion which can be seen from the purely fermionic contribution 
to the probability density which is zero at $r=0$.

Increasing the number of basis functions, the only thing that happens
is that the global maximum moves slowly to larger and larger $r$ at
the same time as the whole wavefunction becomes more spread out in $r$
but more peaked in $\theta$.
One can expect that considering the complete base
(\ref{base}) the ground state density will be the same near the origin as 
in the picture and also will have the hills on the valleys of the 
potential which will be sharper and sharper when we follow the potential 
valley to large $r$. One can not conjecture in this case the behavior of 
the hills by this numerical method. 
Rather one has to use other methods \cite{HS} for the asymptotic behavior 
of the wavefunction at large $r$.

\section{Numerical calculation II}

  Here we briefly review the body of paper \cite{W1} where a different
approach to solving the eigenvalue problem of a Hamiltonian is given.
  The main idea is to choose a finite subset of a complete basis of the
Hamiltonian Hilbert space and 
then find the combination which minimizes the lowest lying
energy for instance. A lot of Hamiltonians have the form (\ref{op})
which can be easily rewritten in terms of the harmonic oscillator
\begin{eqnarray}\label{hamosc}
-\frac{1}{2}\pd{x}+\frac{1}{2}\zeta^2x^2
\end{eqnarray}
creation and annihilation operators $a_+, a_-$. This is true even in
the case where the potential $V(x)$ is not quadratic.
The complete base above is formed by eigenvectors of this linear
harmonic oscillator which are constructed by creation operators acting
on the vacuum. If we choose a finite subset of this basis, calculating
Hamiltonian matrix elements is very easy.

In the harmonic oscillator Hamiltonian (\ref{hamosc}) which we used to
define the basic creation and annihilation operators there is an
arbitrary parameter $\zeta$ in the potential which we may use to
optimize the numerics. The option to use $\zeta$ to optimize the 
calculations was invented by
us and should be regarded as suggestion for an improvement of the
method presented in \cite{W1}. Concretely we will do it like this: first we
choose a subset of the basis functions, we then compute the energy (or
whatever else we would like to calculate) for different values of
$\zeta$. We then find the value of $\zeta$ which gives the
minimum value for the energy. It turns out that it is much more
effective to minimize the energy in terms of $\zeta$ than to try to
use a larger basis since the computer time used for a calculation
increases exponentially with the number of basis functions included.

There is one extra complication in our problem.
Because of the gauge symmetry one should also restrict the basis to
include only gauge invariant states. The generalization to the fermionic 
part and to the higher dimensional problem is straightforward.

  Let us illustrate the method above on our Hamiltonian. First of all we
have to form all gauge invariant states which can be obtained by acting with
a combination of creation operators on the vacuum. The $SU(2)$ 
algebra has only two tensors which can form scalars from vectors
\begin{eqnarray}
\delta^{ab},~\epsilon^{abc} \;.
\end{eqnarray}
These vectors are formed by creation operators ${a_+}_i^a, \chi^a$ and a
combination of them has to be contracted in gauge indexes with the
special tensors above to get a gauge scalar. Let us write all non zero
fundamental possibilities
\begin{eqnarray}
\delta^{ab}{a_+}_i^a{a_+}_j^b\nonumber\\
\delta^{ab}{a_+}_i^a\chi^b\nonumber\\
\epsilon^{abc}{a_+}_i^a{a_+}_j^b\chi^c\nonumber\\
\epsilon^{abc}{a_+}_i^a\chi^b\chi^c\nonumber\\
\epsilon^{abc}\chi^a\chi^b\chi^c
\end{eqnarray}
where any combination of them acting on a gauge invariant state gives
also a gauge invariant state. One has to start with the
vacuum to generate all gauge invariant state. However, if we are
looking for the groundstate we
are interested only in states which can form a state with total spin zero.
That is only states which consist of the fermion vacuum or two creation fermion
operators acting on it because these are the only cases where the
fermionic contribution to the total angular momentum can be canceled
by the bosonic angular momentum. So using the operators
\begin{eqnarray}
{a_+}_i^a{a_+}_j^a \qquad i,j=1,2
\label{boscom}
\end{eqnarray}
acting on $\ket{0}$ and on the fermion states
\begin{eqnarray}
\ket{1'}=\frac{1}{\sqrt{12}}{a_+}_1^a\epsilon^{abc}\chi^b\chi^c\ket{0}
\nonumber\\
\ket{2'}=\frac{1}{\sqrt{12}}{a_+}_2^a\epsilon^{abc}\chi^b\chi^c\ket{0}
\label{fercom}
\end{eqnarray}
gives the appropriate complete basis for our Hamiltonian. Choosing a
finite subset of the basis states
allows us to calculate matrix elements of any operator on the
computer. It is a
good idea to orthonormalize the states in the basis to avoid problems in
the eigenvalue calculation when the metric is non trivial. See the next
section for the explicit form of first few states in the basis.

\begin{figure}[htb]
\begin{center}
\mbox{\epsfxsize=14.2cm\epsfbox{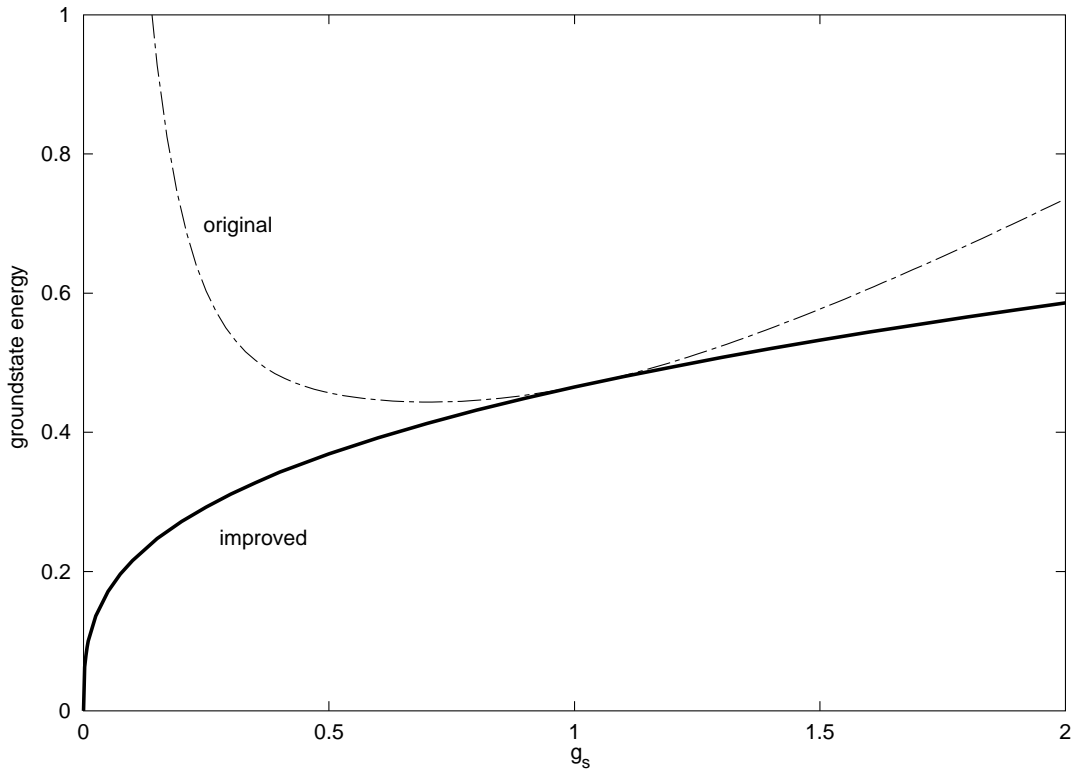}}
\end{center}
\caption{Groundstate energy dependence on the $g_s$}
\label{zeta}
\end{figure}

 We are able to illustrate the effect of the optimized $\zeta$ now. Let us
show what happens with the ground state energy when we calculate with
fixed $\zeta$ but with different coupling constants $g_s\in(0,2]$ in our
Hamiltonian (\ref{toy}). In this calculation we have chosen the
restriction of the basis to include  up to six
creation operators in the bosonic basis and up to five creation
operators in the fermion part. The result is given by the dotted line
in the figure \ref{zeta}.

It seems that the energy of the ground state {\em diverges} when the
coupling constant goes to zero which is clearly wrong! The way out of
this dilemma is to remember that we also have the parameter $\zeta$ at
our disposal. If we for each $g_s$ use the $\zeta$ which gives the
minimal energy, we get the dependence of the ground state energy on
the coupling constant as drawn in bold line. This is
clearly consistent with the $g_s^{1/3}$ dependence one expects
from dimensional arguments.
This is a very nice illustration of the power of our suggestion for
improving the Wosiek method \cite{W1} by including the $\zeta$
parameter. One can guess that there has to exist a coupling constant such
that the optimalization is not needed. This is also viewable from the
picture.

\begin{figure}[htb]
\begin{center}
\mbox{\epsfbox{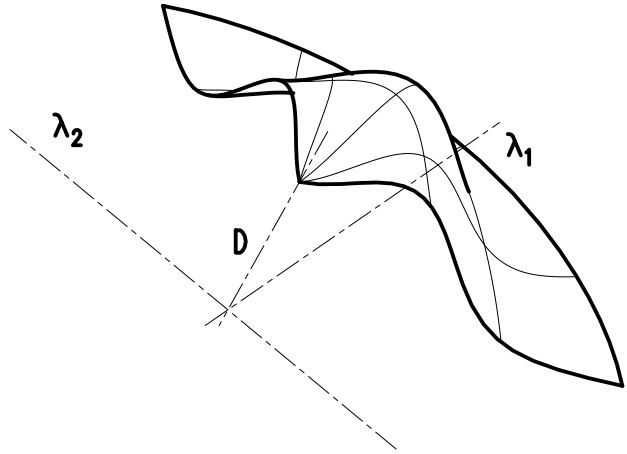}}
\end{center}
\caption{Groundstate by Wosiek method}
\label{figw}
\end{figure}

  There are also some results for coupling constant $g_s=0.1$ in the
(Tab.~\ref{tab}, p. \pageref{tab}) where the meaning of the columns is described above the
table. The energy of the bold line for $g_s=0.1$ corresponds to the
first two columns in the table for the highest order obtained by the
Wosiek method. We draw the
density plot (Fig. \ref{figw}.) for this calculation on which we can see the bump which for
a larger number of basis functions develops into the long and thin hill in
the picture on page \pageref{picden}.
Its domain is $(r,\theta)\in[0,1.5]\times[-\pi/4,\pi/4]$.

\section{Comparison}
\label{comparison}

  To compare our method with the Wosiek method it is important to know the
correspondence between the base (\ref{boson},\ref{fermion}) and the base
formed by creation operators. Since we have written our Hamiltonian in
terms of angular momentum operators it is useful to classify all states
in terms of angular momenta. In fact, this is even more useful since
when we are discussing the ground state with total spin zero.
We may rewrite
the angular momentum operator (\ref{momentum}) as
\begin{eqnarray}
\mathrm{i}\left({a_+}_2^a{a_-}_1^a-{a_+}_1^a{a_-}_2^a\right)
\end{eqnarray}
using the correspondence between creation/annihilation and $X_i^a,\pi_i^a$
operators
\begin{eqnarray}
{a_\pm}_i^a=\sqrt{\zeta/2}(X_i^a\mp\mathrm{i}/\zeta\pi_i^a)\;.
\end{eqnarray}
Let us write down how the lowest lying angular momentum states look
like. First we do the purely bosonic sector. The (zero angular
momentum) ground state $|0\rangle$ in the coordinate 
representation looks like
\begin{eqnarray}
\psi_0(r)=\left(\frac{\zeta}{\pi}\right)^\frac{3}{2}e^{-\frac{1}{2}\zeta r^2}.
\end{eqnarray}
Next we can write down all gauge invariant states created by two
bosonic oscillators.
\begin{eqnarray}
\ket{1'}&=&\frac{1}{\sqrt{6}}{a_+}_1^a{a_+}_1^a\ket{0} \nonumber\\
\ket{2'}&=&\frac{1}{\sqrt{6}}{a_+}_2^a{a_+}_2^a\ket{0} \nonumber\\
\ket{3'}&=&\frac{1}{\sqrt{3}}{a_+}_1^a{a_+}_2^a\ket{0}\;.
\label{first}
\end{eqnarray}
The following linear combinations
\begin{eqnarray}
\frac{\mathrm{i}}{2}\ket{1'}-\frac{\mathrm{i}}{2}\ket{2'}+
\frac{1}{\sqrt{2}}\ket{3'} \nonumber\\
\frac{1}{\sqrt{2}}\ket{1'}+\frac{1}{\sqrt{2}}\ket{2'} \nonumber\\
-\frac{\mathrm{i}}{2}\ket{1'}+\frac{\mathrm{i}}{2}\ket{2'}+
\frac{1}{\sqrt{2}}\ket{3'} 
\end{eqnarray}
are orthonormal eigenvectors of the angular momentum operator with
eigenvalues $\{-2,0,2\}$. 
Using formulas for the creation/annihilation operators in the $X_i^a,
\pi_i^a$ representation we may write the coordinate representation of
these states as
\begin{eqnarray}
\frac{\mathrm{i}}{\sqrt{6}}r^2\zeta\cos 2\theta
            e^{-2\mathrm{i}\phi}\psi_0 \nonumber\\
\frac{1}{\sqrt{3}}(-3+r^2\zeta)\psi_0 \nonumber\\
-\frac{\mathrm{i}}{\sqrt{6}}r^2\zeta\cos 2\theta
            e^{2\mathrm{i}\phi}\psi_0\;.    
\end{eqnarray}
The angular momenta spectrum of the next six states all formed by four
creation operators is $\{-4,-2,0,2,4\}\oplus\{0\}$ where the two
orthonormal states with zero angular momenta are given by
\begin{eqnarray}
\frac{1}{4\sqrt{105}}(60-40r^2\zeta+5r^4\zeta^2
   +3r^4\zeta^2\cos 4\theta)\psi_0 \nonumber\\
\frac{1}{4\sqrt{21}}(36-24r^2\zeta+3r^4\zeta^2
   -r^4\zeta^2\cos 4\theta)\psi_0\;.
\label{fixed}   
\end{eqnarray}
They give the first non trivial wave function dependence on $\theta$ in 
the zero angular momentum sector. The $\theta$ part in
higher order states with zero angular momenta reproduce the basis (\ref{boson})
which we guessed as the Fourier basis for the functions with the correct 
boundary conditions. We see that a new $\cos4j\theta$ appears when
there appears a new ${0}$ in the angular momenta spectrum. 

For reference we also give the
spectra for the states created by six and eight creation operators
$\{-6,-4,-2,0,2,4,6\}\oplus\{-2,0,2\}$ and  
$\{-8,-6,-4,-2,0,2,4,\\6,8\}\oplus\{-4,-2,0,2,4\}\oplus\{0\}$. With this we 
close the discussion about the purely bosonic part.

In the same way we can study the fermionic part.
The wave function that interest us most are the ones with angular
momentum one since they are the ones which contribute to the ground state
with total angular momentum zero.
On the lowest level in the creation/annihilation operator basis we have
two gauge invariant orthonormal states
\begin{eqnarray}
\ket{1'}=\frac{1}{\sqrt{12}}{a_+}_1^a\epsilon^{abc}\chi^b\chi^c\ket{0}
\nonumber\\
\ket{2'}=\frac{1}{\sqrt{12}}{a_+}_2^a\epsilon^{abc}\chi^b\chi^c\ket{0}\;.
\label{second}
\end{eqnarray}
They can be combined into the orthonormal states
\begin{eqnarray}
\frac{\mathrm{i}}{\sqrt{2}}\ket{1'}+\frac{1}{\sqrt{2}}\ket{2'}\nonumber\\
-\frac{\mathrm{i}}{\sqrt{2}}\ket{1'}+\frac{1}{\sqrt{2}}\ket{2'}
\end{eqnarray}
which are eigenstates of angular momenta with eigenvalues $\{-1,1\}$.
It is again possible to rewrite these states in the coordinate basis
with help of the orthonormal and gauge invariant purely fermionic 
states (\ref{states}) as
\begin{eqnarray}
\frac{i}{\sqrt{3}}r\sqrt\zeta\cos\theta e^{-\mathrm{i}\phi}\ket{1}+
\frac{1}{\sqrt{3}}r\sqrt\zeta\sin\theta e^{-\mathrm{i}\phi}\ket{2}
\nonumber\\
-\frac{i}{\sqrt{3}}r\sqrt\zeta\cos\theta e^{\mathrm{i}\phi}\ket{1}+
\frac{1}{\sqrt{3}}r\sqrt\zeta\sin\theta e^{\mathrm{i}\phi}\ket{2}.
\end{eqnarray}
The $\theta$ part of the second state gives the first function in the
basis (\ref{fermion}). The states in the next level
consisting of the six states
we get by acting with all gauge invariant combinations of two bosonic
creation operators on the previous two states have angular momentum
spectrum $\{-3,-1,1,3\}\oplus\{-1,1\}$ where 
the two orthonormal states with angular momentum one are given by
\begin{eqnarray}
\frac{1}{\sqrt{18}}r\sqrt\zeta\cos\theta(2-r^2\zeta+r^2\zeta\cos 2\theta)
    e^{\mathrm{i}\phi}\ket{1}\qquad\qquad \nonumber\\
-\frac{\mathrm{i}}{\sqrt{18}}r\sqrt\zeta\sin\theta
(-2+r^2\zeta+r^2\zeta\cos 2\theta)
    e^{\mathrm{i}\phi}\ket{2}\nonumber\\
-\frac{i}{\sqrt{90}}r\sqrt\zeta\cos\theta
(-10+2r^2\zeta+r^2\zeta\cos 2\theta)
    e^{\mathrm{i}\phi}\ket{1}\qquad\qquad \nonumber\\
-\frac{\mathrm{1}}{\sqrt{90}}r\sqrt\zeta\sin\theta
(10-2r^2\zeta+r^2\zeta\cos 2\theta)
    e^{\mathrm{i}\phi}\ket{2} \;.
\end{eqnarray}
This agrees with the second function in our fermionic basis. Continuing we 
find that the $\theta$ part of the angular momentum one states at higher
levels reproduce the basis (\ref{fermion}) 
which we have used. Writing the angular momentum spectrum for the next
level (states obtained by acting with all gauge invariant combinations
of four bosonic oscillators on the original fermionic states) we have
$\{-5,-3,-1,1,3,5\}\oplus\{-3,-1,1,3\}\oplus\{-1,1\}$. This closes
the discussion about the fermionic part and again gives a hint how
the general spectrum looks like.

We have now achieved our main goal to rewrite the states in the Wosiek
method in terms of angular momentum eigenstates in order to be able to
compare the two methods. Since we now know at what level a particular
basis function appears in the Wosiek method, we know where we should
``cut off'' the set of basis functions used in our method in order to
get an equivalent calculation. Furthermore, doing the calculation
using the two different methods at the same level we should still
expect that our method should give better results since in our method
the radial part of the wave function is not a priori fixed while in
the Wosiek method the radial part comes together with the angular
part, see for instance (\ref{fixed}). Also, our method should be more
effective for fixed spin calculation on the computer since
we have the possibility to use only the angular momentum eigenstates
we need while in the Wosiek method one is using all possible states which
leads to a huge number of states not contributing to the wavefunction but
absorbing space and time in the computer.

To illustrate this we compose  the following table.
\begin{table}[htb]
\begin{center}
\begin{tabular}{|c|c|c|c|c|c|}
\hline
$N_p$ & $\zeta$ & $E_p$ & $N$ & $E$\\
\hline
\hline
4+2        & 4.55 & 0.395 & 1+1   & 0.376 \\
\hline
10+8       & 4.55 & 0.303 & 2+2   & 0.295 \\
\hline
20+20      & 4.73 & 0.216 & 2+3   & 0.214 \\
\hline
1540+440   &      &       & 10+10 & 0.075 \\
\hline
11480+3080 &      &       & 20+20 & 0.037 \\
\hline
37820+9920 &      &       & 30+30 & 0.025 \\
\hline
           &      &       & 40+40 & 0.018 \\
\hline
           &      &       & 50+50 & 0.014 \\
\hline
           &      &       & 80+80 & 0.008 \\
\hline
\end{tabular}
\end{center}
\caption{Results}
\label{tab}
\end{table}
We compare the (improved) Wosiek method with $N_p$ boson and 
fermion creation/annihi\-lation states of the type
(\ref{first},\ref{second}) with our method using only the test
functions (\ref{boson},\ref{fermion}) to the corresponding level  for $g_s=0.1$. The
$\zeta$ given in the table is the optimized value 
for the (improved) Wosiek method, the $N$  counts the number of the
test functions corresponding to the level used in the Wosiek method and 
the groundstate energies obtained by both methods are also given.

We see that indeed the energy calculated using our method is lower
than in the Wosiek method. Furthermore it is clear that the number of
states used in the Wosiek method increases uncontrollably
as compared to the increase of basis functions used in our method.

Finally we
also compare probability density sections for $\theta=0$ in the case
of the third line in the table (the last case where both methods have
been used) on the domain $r\in[0,2]$. 
The states which generate these sections have the same norm to compare them
correctly. 
\begin{figure}[htb]
\begin{center}
\mbox{\epsfxsize=14.2cm\epsfbox{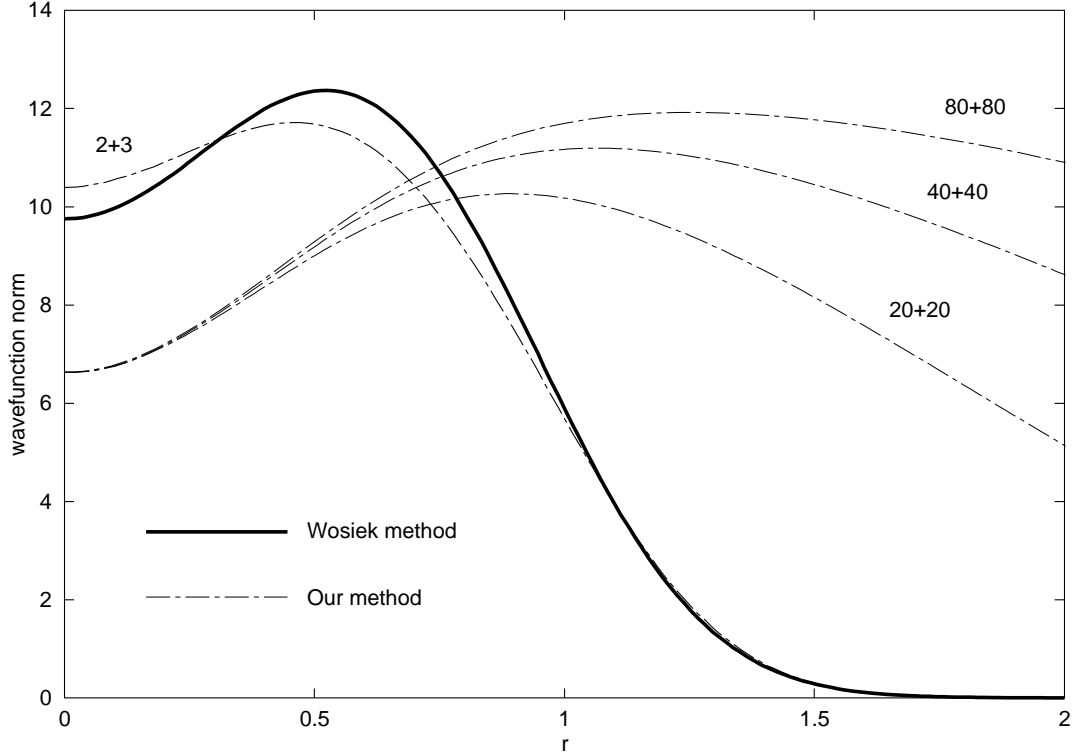}}
\end{center}
\caption{Comparison of wavefuctions at $\theta=0$}
\end{figure}
We have also included higher order wave functions calculated using our
method. The labels indicate which order they correspond to. 
These wave functions are not equally normalized but have rather been
chosen to have the same value at $r=0$ in order to illustrate the
behavior of the wavefunction when the order is increased. We see that
the maximum moves outwards at the same time as the wavefunction
becomes wider.

\section{Conclusions}

  The aim of the present work was to find an approach how to calculate the
boundstate of two 0-branes. This we have achieved with good results. 
In the course of our work we also found a way
to improve the Wosiek method  but our method still gives
better results. As this work was being completed the closely
related paper \cite{W2} by Wosiek (which improve his original method)
appeared.

We have also calculated the groundstate probability
density near the origin with high accuracy which gives a basic
intuition about the physics of branes on the string scale. In particular
about the meaning of the auxilliary coordinates which become important
at small distances between the branes. It is for example interesting
to observe that the most probable position of the branes is not on top
of each other but rather at some small distance away from each
other. This we understand as an effect of the Pauli fermionic repulsion.

It is possible to apply the method described here for branes in
higher dimensional 
Minkowski space. However there are a lot of complications which do not allow
numerical results of high precision. Namely, there are many more
auxilliary coordinates in the higher dimensional case (or in the case
where there are more than two branes).
Since our basis functions are functions of the auxilliary coordinates,
this implies that our basis functions will be multi dimensional
functions and thus the number of components that the computer has to
handle increases. Naively speaking the number of effective radial wavefunction
in the state computed by the Numerov method for a given dimension is
the number in our toy model raised to the power of the number of 
auxilliary coordinates. This fact of course taxes the computer very 
heavily.
Another important point that we wanted to study was how the branes
behave in external (generalized electric-magnetic) fields. The potential  
coming from these background
fields represents this interaction. However, our two dimensional
toy model is too simple to display any interesting effects. Again one
would have to go to higher dimensional Minkowski space to do be able
to do some interesting calculations.

As a first attempt to go further it would therefore be nice to
consider three dimensional Minkowski space which would allow
more general background fields which coming from other types of
branes. However, we do feel that doing {\em realistic} calculations of
the full ten dimensional (or even four dimensional) theory or
including more than two 0-branes would not
be possible even on a super computer. This is one of the reasons why
we have not pursued this issue further.

\section*{Acknowledgments}

I am very grateful to Rikard von Unge for continuous support and I also
thank to Milan \v{S}indelka which told me of his experiences in numerical
methods which can solve quantum mechanics problems.

\appendix

\section{Metric derivation}
\label{appmetric}

\def\theequation{\thesection.\arabic{equation}}
\setcounter{equation}{0}

It is possible to decompose $X_i^{~a}$ \cite{HS} in matrix form as
(\ref{decomposition})
\begin{eqnarray}
X=\psi\Lambda\eta
\end{eqnarray}
where $\psi$ is a group element in the adjoint representation of the gauge 
group and $\eta$ is a group element of the rotation group $SO(2)$.
The matrix $\Lambda$ is then gauge and rotation invariant. We may now 
choose our physical coordinates as
\begin{eqnarray}
\Lambda&=&\left(\matrix{\lambda_1 & 0 \cr 0 & \lambda_2  \cr
0 &0}\right),~~\lambda_1=r\cos\theta;\;\lambda_2=r\sin\theta
\label{rt}
\end{eqnarray}
together with
\begin{eqnarray}
\eta=\exp\left(\phi T\right) \;,
\label{eta}
\end{eqnarray}
where $T$ is the generator of spatial rotation
\begin{eqnarray}
\left(\matrix{0 & 1 \cr -1 & 0}\right)\;.
\end{eqnarray}
The matrix $\psi$  which describes the gauge degrees of
freedom is parametrized by Euler angles $\alpha,\beta,\gamma$
\begin{eqnarray}
\psi=\exp\left(\alpha T_3\right) \exp\left(\gamma T_1\right)
     \exp\left(\beta T_3\right)
\label{psi}
\end{eqnarray}
where the matrices are generators of $SU(2)$ in the adjoint 
representation.

$\;$

\noindent
The flat metric in the original $X^{a}_{i}$ coordinates is trivial and
can be written in matrix notation as
\begin{eqnarray}
g=\mathrm{Tr}(dX^TdX) \;.
\end{eqnarray}
The decomposition (\ref{decomposition}) leads to the expression
\begin{eqnarray}
dX=d\psi\Lambda\eta+\psi d\Lambda\eta+\psi\Lambda d\eta
\end{eqnarray}
and then the metric is
\begin{eqnarray}
\mathrm{Tr}(d\Lambda^T d\Lambda-\psi^Td\psi\Lambda\Lambda^T\psi^T
d\psi+d\eta^T\Lambda^T\Lambda d\eta+2\eta d\eta^T\Lambda^T\psi^Td\psi\Lambda) \;.
\label{expanded}
\end{eqnarray}
We see here that the differentials in $d\Lambda$ are not mixed with the
other ones and thus the metric will be block diagonal
\begin{eqnarray}
\left(\matrix{g_i & \cr & g_m}\right)
\label{diag}
\end{eqnarray}
where the $g_i$ block is associated with the $d\Lambda$ differential
and $g_m$ is associated with the differentials $d\eta,d\psi$. Now we
compute the metric term by term. The $\Lambda$  part in the case of two 
branes is
\begin{eqnarray}
\mathrm{Tr}(d\Lambda^Td\Lambda)= d\lambda_i d\lambda_i
\end{eqnarray}
or explicitly in variables $r,\theta$ from (\ref{rt})
\begin{eqnarray}
d\lambda_i d\lambda_i=dr^2+r^2d\theta^2 \;.
\label{firstpart}
\end{eqnarray}
Next we take care of the terms which contribute to the $g_m$ part of the
metric. Using the fact that
\begin{eqnarray}
\psi^{-1}=\psi^T
\end{eqnarray}
the purely gauge part
\begin{eqnarray}
-\mathrm{Tr}(\psi^Td\psi\Lambda\Lambda^T\psi^Td\psi)
\label{gaugepart}
\end{eqnarray}
can be written with help of the left-invariant one forms defined by
the $\psi$ being a group element in the adjoint representation of $SU(2)$
\begin{eqnarray}
\omega_L=\psi^{-1}d\psi=\omega_{La}T_a
\label{leftdef}
\end{eqnarray}
where $T_a$ are the generators. In our coordinates we explicitly have 
the expressions
\begin{eqnarray}
\omega_{L1}&=&\sin\gamma\sin\beta d\alpha+\cos\beta d\gamma \nonumber\\
\omega_{L2}&=&-\sin\gamma\cos\beta d\alpha+\sin\beta d\gamma \nonumber\\
\omega_{L3}&=&\cos\gamma d\alpha+d\beta \;.
\label{leftforms}
\end{eqnarray}
Introducing the matrix $\Omega_L$ of the coefficient of the $\omega$
1-forms
\begin{eqnarray}
\label{omegadef}
\left(\matrix{\omega_{L1}\cr \omega_{L2}\cr \omega_{L3}}\right)=
\underbrace{
\left(
  \matrix{\sin\gamma\sin\beta & 0 & \cos\beta \cr
             -\sin\gamma\cos\beta & 0 & \sin\beta \cr
	     \cos\gamma & 1 & 0}
\right)}_{\displaystyle\Omega_L}
\left(
\matrix{d\alpha\cr d\beta\cr d\gamma}
\right)
\end{eqnarray} 
the gauge part (\ref{gaugepart}) is 
\begin{eqnarray}
e^T\Omega_L^T\Pi\Omega_L e
\label{gauge}
\end{eqnarray}
where $e^T=(d\alpha,d\beta,d\gamma)$ and the diagonal matrix $\Pi_{ab} = 
\mathrm{Tr}(T_a\Lambda\Lambda^T T_b)$ has elements
\begin{eqnarray}
\Pi_{aa}=\sum_{i\neq a}\lambda_i^2 \;.
\end{eqnarray}
The calculation of the contribution from terms which contain $d\eta$
is very similar. Introducing the right-invariant one form defined by
 $d\eta\eta^{-1}$ and using our parametrisation (\ref{eta}) we have
\begin{eqnarray}
\mathrm{Tr}(d\eta^T\Lambda^T\Lambda d\eta)=r^2 d\phi^2
\label{angular}
\end{eqnarray}
and 
\begin{eqnarray}
\mathrm{Tr}(2\eta d\eta^T\Lambda^T\psi^Td\psi\Lambda)
=2r^2 \sin 2\theta d\phi \omega_{L3} \;.
\label{mixing}
\end{eqnarray}
In this expression it is necessary to write the $\omega_L$ (\ref{leftforms})
explicitly. Since we know that the metric is block diagonal (\ref{diag}) 
we will write the $g_m$ part of the metric compactly. This is done 
by writing the matrix (\ref{boldpi})
\begin{eqnarray}
{\bf\Pi}=\left(\matrix{r^2 & & & r^2\sin 2\theta \cr
 & r^2\sin^2\theta & & \cr
&  & r^2\cos^2\theta & \cr
r^2\sin 2\theta & & & r^2}\right)
\end{eqnarray}
and the structure matrix
\begin{eqnarray}
{\bf \Omega}_L=\left(\matrix{1 & \cr & \Omega_L}\right) \;.
\label{omega}
\end{eqnarray}
Then it is possible to write the sum of (\ref{gauge},\ref{angular},
\ref{mixing}) as
\begin{eqnarray}
{\bf e}^T{\bf\Omega}_L^T{\bf\Pi}{\bf\Omega}_L{\bf e}
\end{eqnarray}
where ${\bf e}^T=(d\phi,d\alpha,d\beta,d\gamma)$. So
\begin{eqnarray}
g_m={\bf\Omega}_L^T{\bf\Pi}{\bf\Omega}
\label{egm}
\end{eqnarray}
and the full metric is
\begin{eqnarray}
dr^2+r^2d\theta^2+{\bf
e}^T{\bf\Omega}_L^T{\bf\Pi}{\bf\Omega}_L{\bf e} \;.
\label{metric}
\end{eqnarray}

It is now easy to find the square root of the metric determinant
\begin{eqnarray}
\frac{1}{4}r^5\sin\gamma\sin 4\theta
\label{measure}
\end{eqnarray}
which also gives us the integration measure. On the domain of
$(r,\theta,\phi,\alpha,\beta,\gamma)$
\begin{eqnarray}
(0,\infty)\times[0,\pi/4]\times(0,2\pi]
\times(0,2\pi]\times(0,2\pi]\times(0,\pi]
\label{domain}
\end{eqnarray}
the transformation rule (\ref{decomposition}) is one to one and on its inside
the Jacobian is
nonzero and thus regular which is important in calculation of 
derivatives with respect to $X_i^{~a}$ in terms of these coordinates.

\section{Laplacian derivation}
\label{applaplace}

\def\theequation{\thesection.\arabic{equation}}
\setcounter{equation}{0}

The metric in the new coordinates is (\ref{metric})
\begin{eqnarray}
dr^2+r^2d\theta^2+{\bf
e}^T{\bf\Omega}_L^T{\bf\Pi}{\bf\Omega}_L{\bf e} \;
\end{eqnarray}
where ${\bf e}^T=(d\phi,d\alpha,d\beta,d\gamma)$, ${\bf\Omega}_L$ is
defined by (\ref{omega}) and ${\bf\Pi}$ by (\ref{boldpi}). From the 
metric or directly from (\ref{expanded}) one can see that the metric has 
the block diagonal structure (\ref{diag})
\begin{eqnarray}
\left(\matrix{g_i(r) & \cr & g_m(r,\theta,\alpha,\beta,\gamma)}\right)
\end{eqnarray}
where $g_i$ is a diagonal matrix being the metric in the $r,\theta$ variables
and $g_m$ is the metric in the other variables. Thus we are able to calculate
the inverse of the full metric in terms of the inverses of $g_i$ and $g_m$
separately. Since $g_i$ does not depend on the gauge variables, the 
Laplacian also splits into two pieces
\begin{eqnarray}
\frac{1}{\sqrt {\det g_i\det g_m}}\partial_j(\sqrt {\det g_i\det g_m} 
g_i^{jj}\partial_j)+\frac{1}{\sqrt {\det g_m}}\partial_\mu(\sqrt {\det 
g_m} g_m^{\mu\nu}\partial_\nu)
\label{pieces}
\end{eqnarray}
where the latin index runs over the $\{r,\theta\}$ variables and the 
greek indexes over the angular variables. Using the square root of the 
metric determinant (\ref{measure}) the first part of the Laplacian 
generated by $g_i$ is
\begin{eqnarray}
\frac{1}{\sqrt {\det g}}\partial_j(\sqrt{ \det g} g_i^{jj}\partial_j)=
\frac{1}{r^5}\p{r}r^5\p{r}
+\frac{1}{r^2\sin4\theta}\p{\theta}\sin4\theta\p{\theta} \;.
\end{eqnarray}
At this point
we would like to rewrite the second part of the Laplacian in terms of the 
angular momenta (\ref{momentum},\ref{gaugemomenta})
\begin{eqnarray}
L^0&=&-i\p\phi\nonumber\\
L^1&=&-i\cot\gamma\sin\alpha\p{\alpha}+i\csc\gamma\sin\alpha\p{\beta}
      +i\cos\alpha\p{\gamma} \nonumber \\
L^2&=&-i\cot\gamma\cos\alpha\p{\alpha}+i\csc\gamma\cos\alpha\p{\beta}
      -i\sin\alpha\p{\gamma} \nonumber \\
L^3&=&i\p{\alpha}\;.
\end{eqnarray}
We notice that the angular momenta are equal to the vector field $X_R$ 
being dual to the {\em right} invariant one form $\omega_R = d\psi 
\psi^{-1}$
\begin{eqnarray}
-iL_1=X_{R1}&\equiv&-\cot\gamma\sin\alpha\p{\alpha}+\csc\gamma\sin\alpha\p
{\beta}+\cos\alpha
\p{
\gamma} \nonumber \\
-iL_2=X_{R2}&\equiv&-\cot\gamma\cos\alpha\p{\alpha}+\csc\gamma\cos\alpha\p{\beta}
-\sin\alpha
\p{\gamma} \nonumber \\
-iL_3=X_{R3}&\equiv&\p{\alpha}\;.
\end{eqnarray}
Since the right invariant one forms induce left translations, they are 
Killing fields of the metric.
Also, since we know the relation between the left and right invariant one 
forms
\begin{eqnarray}
\label{LRrel}
\omega_R=d\psi \psi^{-1}=\psi\omega_L\psi^{-1}
\end{eqnarray}
we may transform the expression for the metric, which is expressed in 
terms of $X_L$ into an expression in terms of $X_R$ which is 
then equivalent to an expression in terms of the angular momenta. 
Explicitly we write the vector field $X_{L,R}$ in terms of 
coefficient matrices $\Theta_{L,R}$ inverse to the matrices $\Omega_{L,R}$ 
introduced in (\ref{omegadef}). The relation (\ref{LRrel}) then implies
\begin{eqnarray}
\Theta_L=\psi^{-1}{\Theta}_R \;.
\label{trans}
\end{eqnarray}
To treat all the angles on equal footing, we incorporate also the physical 
angle $\phi$ into this formalism by extending the 
three by three matrices $\Omega_R,\Theta_R,\psi$ to four by four matrices 
by putting $-1$ in the "zeroth" row and column. These new matrices we 
denote by bold letters, for instance
\begin{eqnarray}
{\bf \Psi}=\left(\matrix{-1 & \cr & \psi}\right) \;.
\end{eqnarray}
Now all vector fields obtained from ${\bf \Theta}_R$ are equal
to $-\mathrm{i}$ times the physical and the gauge angular momenta and they 
are all Killing vectors of the metric. We can now write 
$\bf \Omega_L = \Psi^{-1}\Omega_R$; $\bf\Theta_L=\Psi^{-1}\Theta_R$.
 
We may now rewrite the angular part of the Laplacian in terms of the 
angular momenta which are given by $\bf\Theta_R$. The appropriate part in 
the Laplacian is
\begin{eqnarray}
{\bf\Theta}_L^T{\bf\Pi}^{-1}{\bf\Theta}_L\;\;\nonumber\\
={\bf\Theta}_R^T{\bf\Psi}{\bf\Pi}^{-1}{\bf\Psi}^{-1}
{\bf\Theta}_R\;.
\end{eqnarray}
First we expand the formula for the Laplacian
\begin{eqnarray}
\frac{1}{\sqrt g}\partial_\mu({\sqrt g} g^{\mu\nu}\partial_\nu)=
\frac{1}{\sqrt g}\partial_\mu(\sqrt{g})g^{\mu\nu}\partial_\nu+
\partial_\mu (g^{\mu\nu})\partial_\nu+
g^{\mu\nu}\partial_\mu\partial_\nu \;.
\end{eqnarray}
The first term is reduced to
\begin{eqnarray}
-\p{x^\xi}({\bf\Theta}_R)_\sigma^{~\xi}({\bf\Psi}
{\bf\Pi}^{-1}{\bf\Psi}^{-1}{\bf\Theta}_R)_\sigma^{~\rho}\p{x^\rho}
\end{eqnarray}
where we have used the Killing equation for ${\bf\Theta}_R$. In the
same way the second term becomes
\begin{eqnarray}
\p{x^\xi}({\bf\Theta}_R)_\sigma^{~\xi}({\bf\Psi}{\bf\Pi}^{-1}{\bf\Psi}^{-1}
{\bf\Theta}_R)_\sigma^{~\rho}\p{x^\rho}+({\bf\Theta}_R)_\sigma^{~\xi}
({\bf\Psi}{\bf\Pi}^{-1}{\bf\Psi}^{-1})_{\sigma\zeta}\p{x^\xi}
({\bf\Theta}_R)_\zeta^{~\rho}\p{x^\rho}
\end{eqnarray}
and finally the third part is
\begin{eqnarray}
({\bf\Theta}_R)_\sigma^{~\xi}({\bf\Psi}{\bf\Pi}^{-1}
{\bf\Psi}^{-1})_{\sigma\zeta}({\bf\Theta}_R)_\zeta^{~\rho}
\p{x^\xi}\p{x^\rho}
\;.
\end{eqnarray}
So the sum of these three contributions gives
\begin{eqnarray}
({\bf\Theta}_R)_\sigma^{~\xi}({\bf\Psi}{\bf\Pi}^{-1}
{\bf\Psi}^{-1})_{\sigma\zeta}({\bf\Theta}_R^T)_\zeta^{~\rho}\p{x^\xi}
\p{x^\rho}+
({\bf\Theta}_R)_\sigma^{~\xi}({\bf\Psi}{\bf\Pi}^{-1}
{\bf\Psi}^{-1})_{\sigma\zeta}\p{x^\xi}({\bf\Theta}_R)_\zeta^{~\rho}
\p{x^\rho}\nonumber\\
=({\bf\Psi}{\bf\Pi}^{-1}{\bf\Psi}^{-1})_{\sigma\zeta}
({\bf\Theta}_R)_\sigma^{~\xi}\p{x^\xi}\left[({\bf\Theta}_R)_\zeta^{~\rho}
\p{x^\rho}\right]=
-({\bf\Psi\Pi}^{-1}{\bf\Psi}^{-1})_{\mu\nu}L^\mu L^\nu\nonumber\\
\end{eqnarray}
where we have used that the angular momenta are defined in terms of
the matrix ${\bf \Theta}_R$. Then the whole Laplacian is
\begin{eqnarray}
\frac{1}{r^5}\p{r}r^5\p{r}+
\frac{1}{r^2\sin4\theta}\p{\theta}\sin4\theta\p{\theta}
-({\bf\Psi\Pi}^{-1}{\bf\Psi}^{-1})_{\mu\nu}L^\mu L^\nu \;.
\end{eqnarray}

\end{document}